\documentclass{PoS}
\usepackage{amsmath,amssymb}

\def\del{\partial}
\def\ee{{\mathrm{e}}}
\def\alangle{\Big\langle\!\!\Big\langle}
\def\arangle{\Big\rangle\!\!\Big\rangle}
\def\m{\tilde{m}}
\def\beq{\begin{equation}}
\def\eeq{\end{equation}}
\renewcommand{\epsilon}{\varepsilon}
\renewcommand{\bar}{\overline}
\newcommand{\Sp}{\text{Sp}}
\newcommand{\SU}{\text{SU}}
\newcommand{\U}{\text{U}}
\newcommand{\Z}{\mathbb{Z}}
\newcommand{\1}{\mathbf{1}}
\newcommand{\D}{\mathcal{D}}
\DeclareMathOperator{\re}{Re}
\DeclareMathOperator{\tr}{Tr}
\DeclareMathOperator{\pf}{Pf}

\title{Chiral Lagrangian and spectral sum rules for two-color QCD at
  high density}

\ShortTitle{Chiral Lagrangian and spectral sum rules for two-color QCD
  at high density}

\author{\speaker{Takuya Kanazawa}\\
	Department of Physics, The University of Tokyo, Tokyo 113-0033, Japan\\
	E-mail: \email{tkanazawa@nt.phys.s.u-tokyo.ac.jp}}
\author{Tilo Wettig\\
	Department of Physics, University of Regensburg, 93040 Regensburg, Germany\\
	E-mail: \email{tilo.wettig@physik.uni-regensburg.de}}
\author{Naoki Yamamoto\\
	Department of Physics, The University of Tokyo, Tokyo 113-0033, Japan\\
	E-mail: \email{yamamoto@nt.phys.s.u-tokyo.ac.jp}}

\abstract{%
We report on our analytical study of two-color QCD with an even number of
  flavors at high baryon density. Based on the pattern of chiral symmetry breaking
  induced by BCS-type diquark pairing
  we construct the low-energy effective Lagrangian for 
  the Nambu-Goldstone bosons.
  We also identify a new epsilon-regime at high baryon density and derive 
  Leutwyler-Smilga-type spectral sum rules for the complex eigenvalues
  of the Dirac operator in terms of the fermion gap.
  Our results can in principle be tested in lattice QCD simulations.
}

\FullConference{The XXVII International Symposium on Lattice Field Theory - LAT2009\\
		 July 26-31 2009\\
		 Peking University, Beijing, China}

\begin{document}

\section{Introduction}

Understanding the phase structure of Quantum Chromodynamics (QCD) at 
nonzero temperature ($T$) and quark chemical potential ($\mu$) is an important 
subject relevant to many areas of physics, including relativistic 
heavy ion collisions, the early universe, and the interior of neutron 
stars \cite{Rajagopal2000,Alford2008}. 
Although QCD at $T>0$ and $\mu=0$ has been extensively investigated by 
both analytical methods and first-principle lattice simulations, 
QCD at $\mu\ne 0$ is much less understood, as the
fermion sign problem makes Monte Carlo simulations extremely difficult. 
One exception is QCD in the high-density limit ($\mu\gg\Lambda_{\rm QCD}$) 
where the QCD coupling is weak: A color superconductor called
color-flavor locked (CFL) phase was shown to be realized \cite{Alford:1998mk}
in which color and flavor symmetries lock each other and chiral symmetry 
is spontaneously broken.

At $\mu=0$ an intimate relation has been established between the 
spectral properties of the Dirac operator and the spontaneous breaking 
of chiral symmetry. 
In 1992 Leutwyler and Smilga \cite{Leutwyler1992} 
succeeded in deriving details of the Dirac eigenvalue distribution 
from the low-energy effective theory of QCD in a finite volume 
(the $\epsilon$-regime)
and found that the distribution of Dirac eigenvalues near zero is governed 
by the nonvanishing chiral condensate. Soon after their work, 
Verbaarschot et al.\ discovered that chiral random matrix theory (ChRMT), 
which possesses the same global symmetries as QCD, provides rich information 
on spectral correlations of the Dirac operator 
\cite{Shuryak:1992pi,Verbaarschot:2000dy}, 
leading to the realization that the way in which the thermodynamic limit of the 
spectral density near zero is approached is universal.
This feature enables us to determine the magnitude of 
the chiral condensate in the QCD vacuum with considerable accuracy
by matching the numerical results from lattice QCD simulations against the exact 
analytical results obtained in ChRMT \cite{Fukaya:2007fb}. 
In the past few years considerable progress has also been made in
ChRMT at $\mu\ne0$ (see \cite{Splittorff:2008sw}
for a review), but the high-density region $\mu\gg\Lambda_{\rm QCD}$
is largely unexplored so far.

Recently two of us pointed out a new $\epsilon$-regime specific for the
BCS-state of QCD at high density, where exact analytical 
results, including Leutwyler-Smilga-type spectral sum rules characterized 
by the fermion gap $\Delta$, can be derived \cite{Yamamoto:2009ey}. 
This explicitly shows that the Dirac spectrum at high density is governed 
by the BCS gap but not by the chiral condensate. 
In this report we apply these lines of analysis to two-color QCD 
with an even number of flavors. Our principal motivation comes from the fact 
that two-color QCD can be simulated on the lattice even at $\mu\ne 0$. 
Although this property makes the theory a very 
attractive testing ground for methods and concepts developed 
in the studies of three-color dense quark matter,
the related works have focused on the Bose-Einstein
condensate of the diquarks (see, e.g., \cite{Kogut2000}).
On the other hand, in this report we will concentrate on the 
BCS superfluid, which is the genuine two-color counterpart of 
the color superconductivity (e.g., CFL phase) in the three-color case.
Testing our results on the lattice will provide the first 
signature of BCS pairing as well as determine the value of the gap.

\section{Low-energy effective theory}   \label{2}

Let us first construct the low-energy effective theory for dense two-color QCD \cite{Kanazawa:2009ks}. 
We first define our notation. The fermionic part of the Lagrangian in Euclidean
space reads $\bar\psi(\D(\mu)+\mathcal{M})\psi$ with the $\mu$-dependent Dirac operator 
\beq
  \D(\mu)=\gamma_\nu D_\nu+\gamma_0\mu
\eeq
and the mass term
\beq
  \mathcal{M}=\frac{1}{2}(1+\gamma_5)M+\frac{1}{2}(1-\gamma_5)M^\dagger\,.
\eeq
Here, $\psi$ is a short-hand notation for $N_f$ flavors of two-color
Dirac spinor fields transforming in the fundamental representation of
$\SU(2)_\text{color}$.  The $\gamma_\nu$ are hermitian $\gamma$-matrices. 
The covariant derivative $D_\nu$ is an anti-hermitian operator so that
the eigenvalues of $\gamma_\nu D_\nu$ are purely imaginary. $M$ is the
$N_f\times N_f$ quark mass matrix.
For $M=0$ and $\mu=0$, the fermionic part of the Lagrangian is symmetric under $\U(2N_f)$. For $\mu\ne0$, 
this symmetry is broken explicitly to $\SU(N_f)_L\times\SU(N_f)_R\times \U(1)_B\times \U(1)_A$.
A remarkable property of two-color QCD is that the
fermion sign problem is absent at nonzero $\mu$:
Because of the pseudo-reality of $\SU(2)$ we have $\D(\mu)\tau_2 C
\gamma_5=\tau_2 C \gamma_5 \D(\mu)^{*}$, where $C$ is the charge
conjugation operator and $\tau_2$ is a generator of
$\SU(2)_\text{color}$.  Together with chiral symmetry, $\{\gamma_5,
\D(\mu)\}=0$, it follows that if $\lambda$ is one of the
eigenvalues of $\D(\mu)$, so are $-\lambda,\,\lambda^*$, and
$-\lambda^*$. Consequently, the fermion
determinant is real and non-negative in two-color QCD with an even number of flavors with pairwise degenerate quark masses.

At $\mu \gg \Lambda_{\rm QCD}$, 
perturbative one-gluon exchange indicates that the color antisymmetric channel $\mathbf{1}$ (coming from 
$\mathbf{2}\otimes\mathbf{2}=\mathbf{3}\oplus \mathbf{1}$) is attractive, which implies that the Fermi 
surface becomes unstable and subject to the formation of Cooper pairs, as known in the BCS theory. As a result 
a gap $\Delta$ appears in the spectrum of quasiquarks near the Fermi surface. 
The Pauli principle forces the condensation to occur in the
flavor-antisymmetric channel: 
\beq
  0\ne \langle \psi\psi\rangle\equiv
  \langle \epsilon_{ab}(\psi_a^T)^i C\gamma_5 I^{ij}\psi_b^j \rangle\,,
  \label{eq:diquark}
\eeq
where $a,b\in\{1,2\}$ and $i,j\in\{1,\dots,N_f\}$ are color and 
flavor indices, respectively.  The $N_f\times N_f$ symplectic matrix $I$ is defined as
\beq
  I=\begin{pmatrix}0&-\1\\\1&0\end{pmatrix},
\eeq
where $\1$ is the $(N_f/2)\times(N_f/2)$ unit matrix. Hereafter we always assume 
that $N_f$ is even.

Some comments are in order.
Although the diquark pairing (\ref{eq:diquark}) has the same quantum numbers as
that in \cite{Kogut2000}, their physical meanings are different:
The pairing (\ref{eq:diquark}) is a weakly-coupled BCS-type condensate whereas
that in \cite{Kogut2000} is a strongly-coupled Bose-Einstein condensate (BEC),
and there could be a smooth crossover between the two. 
The crossover from low to high densities, if realized, is a typical BEC-BCS 
crossover known in condensed matter physics.
It parallels the idea of quark-hadron continuity \cite{Schafer:1998ef} in real (three-color) QCD,
which may be explicitly realized by the effect of the axial anomaly \cite{Hatsuda:2006ps}.  
However, for two-color QCD with even $N_f$, the axial anomaly
never acts as an external field for the chiral condensate, and a new mechanism is necessary to account for the 
crossover phenomenon.

The condensation (\ref{eq:diquark}) breaks chiral symmetry spontaneously as follows:
\beq
  \SU(N_f)_L\times\SU(N_f)_R\times \U(1)_B\times \U(1)_A  \to  \Sp(N_f)_L\times \Sp(N_f)_R.
  \label{pattern}
\eeq
We used the fact that the $\U(1)_A$-anomaly is suppressed at high density. It is a characteristic of 
two-color QCD that Cooper pairs are color singlets and preserve gauge symmetry; the system is in the superfluid phase but
not in the superconducting phase. 
We stress that (\ref{pattern}) is different from the way chiral symmetry is spontaneously broken at 
$\mu=0$ \cite{Kogut2000}: $\SU(2N_f)\to \Sp(2N_f)$. (Thus the resulting effective chiral Lagrangian is entirely different as well.) 
The Nambu-Goldstone (NG) fields associated with (\ref{pattern}) are gapless in the chiral limit and govern the low-energy physics 
near the Fermi surface. Let us label the NG modes as
\beq
  \Sigma_L\in \SU(N_f)_L/\Sp(N_f)_L\,,\quad \Sigma_R\in \SU(N_f)_R/\Sp(N_f)_R\,,\quad V\in \U(1)_B\,,\quad A\in \U(1)_A\,.
\eeq
For $N_f=2$, $\Sigma_{L,R}$ do not exist since $\SU(2)\simeq\Sp(2)$. First consider $N_f\geq 4$. In the presence of 
explicit breaking of chiral symmetry by $\mathcal M$, the above NG modes acquire a small but nonzero mass. From symmetry 
arguments plus weak-coupling calculations, the effective chiral Lagrangian valid at energy scales below $\Delta$ is determined to be
\begin{align}
  \mathcal{L}=\ &\frac{f_{H}^2}{2}\Big\{|\del_0 V|^2-v_{H}^2|\del_i
  V|^2\Big\} + \frac{N_ff_{\eta'}^2}{2}\Big\{|\del_0
  A|^2-v_{\eta'}^2|\del_i A|^2\Big\} 
  \notag\\
  &+\frac{f_{\pi}^2}{2}\tr\Big\{|\del_0\Sigma_L|^2-v_{\pi}^2|
  \del_i\Sigma_L|^2+(L\leftrightarrow R)\Big\} -\frac{3\Delta^2}{4\pi^2}
  \Big\{A^2\tr(M\Sigma_R M^T\Sigma_L^\dagger)+\text{c.c.}\Big\}\,.
\end{align}
The $f$'s are decay constants for each NG mode, and the $v$'s are the corresponding velocities originating from the absence of Lorentz
invariance in the medium. The absence of an $O(M)$ term in the chiral Lagrangian is a consequence of the $\Z(2)_L \times
\Z(2)_R$ symmetry of the diquark pairing.

For $N_f=2$, a similar analysis leads to
\beq
\mathcal{L}=\frac{f_{H}^2}{2}\Big\{|\del_0 V|^2-v_{H}^2|\del_i
  V|^2\Big\} + f_{\eta'}^2\Big\{|\del_0
  A|^2-v_{\eta'}^2|\del_i A|^2\Big\}
  -\frac{3\Delta^2}{2\pi^2}\big\{(\det M)A^2+\text{c.c.}\big\}\,.
\eeq
As $V$ and the gluon fields decouple from the other NG modes, they will be neglected in the following.
We note that the chiral Lagrangian at large $\mu$ presented above is a new result.
The Lagrangian in \cite{Kogut2000} is valid only for densities
corresponding to $\mu<m_{\rho}$, where $m_{\rho}$ is the $\rho$-meson mass.

\section{Partition function in a finite volume}   \label{3}

Next we show that the $\epsilon$-regime introduced at $\mu=0$ in \cite{Gasser:1987ah} can be defined at large $\mu$ as well. 
Let us take the imaginary-time formalism and consider two-color QCD in
a finite box of size $L^4\,(\equiv V_4)$. The masses of the 
NG modes due to the quark mass matrix $M$ are denoted by $m_\text{NG}$.
The point is that the dynamics of the system simplifies drastically 
in the regime
\beq
  \frac{1}{\Delta}\ll L\ll \frac{1}{m_\text{NG}}\,.
  \label{eq:regime}
\eeq
The first inequality guarantees that contributions of non-NG modes (e.g., quarks) to the partition function $Z$ are sufficiently 
small: $\mathrm{e}^{-L\Delta}\ll 1$. The second inequality implies that the Compton wavelength of the NG modes is much larger than the linear 
extent of the box, which allows us to truncate the Hilbert space of the NG modes to its zero-momentum sector alone. 
Therefore the partition function in the $\epsilon$-regime (\ref{eq:regime}) is given simply by
\beq
  Z(M)
  =\int\limits_{\U(1)_A}\hspace{-1mm}dA\hspace*{-6mm}
  \int\limits_{\hspace*{5mm}\SU(N_f)_L/\Sp(N_f)_L}\hspace{-10mm}
  d\Sigma_L\hspace{1mm}
  \int\limits_{\hspace*{5mm}\SU(N_f)_R/\Sp(N_f)_R}\hspace{-10mm} 
  d\Sigma_R\hspace{3mm}
  \exp\left[V_4 \frac{3\Delta^2}{4\pi^2} \Big\{A^2\tr(M\Sigma_R
  M^T\Sigma_L^\dagger)+\text{c.c.}\Big\}\right],
  \label{eq:Z_macro}
\eeq
normalized so that $Z(0)=1$. The infinite-dimensional path integral is now reduced to ordinary integrals. 
For degenerate masses ($M=m\mathbf{1}$), the integration can be carried out explicitly, yielding
\beq
  Z(m)=\frac{1}{(N_f-1)!!}\pf(A)\,, 
  \label{eq:Z_exact}
\eeq
where $A$ is an $N_f\times N_f$ antisymmetric matrix with entries
\beq
  A_{pq}\equiv (q-p)I_{p+q}\left(\frac{3}{\pi^2} V_4 \Delta^2 m^2\right)\,,\qquad
  p,q=-\frac{N_f-1}{2},\ldots, \frac{N_f-3}{2},\frac{N_f-1}{2}
  \label{eq:A}
\eeq
and $I_{p+q}$ denotes a modified Bessel function. It is intriguing that $Z$ in (\ref{eq:Z_exact}) bears strong resemblance to 
the known expression for the finite-volume partition function at $\mu=0$
in the topologically trivial sector \cite[(5.13)]{Smilga1995}.  The latter is obtained by 
changing $N_f$ and $\displaystyle (3/\pi^2) V_4 \Delta^2 m^2$ in (\ref{eq:Z_exact}) and (\ref{eq:A}) 
to $2N_f$ and $V_4\Sigma m$, respectively, with $\Sigma$ the magnitude of the chiral condensate. 
It would be interesting to generalize our results to intermediate densities, interpolating between both results.

For $N_f=2$, explicit integration is possible for arbitrary $M$, with the result
\beq
  Z(M)=I_0\left(\frac{3}{\pi^2}V_4\Delta^2\det M\right).
  \label{eq:Z_macro_2}
\eeq

\section{Spectral sum rules for the Dirac operator}   \label{4}

In this section we review the derivation of spectral sum rules and briefly discuss their physical implications. 
Let us denote the complex eigenvalues of the Dirac operator $\D(\mu)$ by $i\lambda_n$, where 
the $\lambda_n$ are real for $\mu=0$. 
Starting from the microscopic Lagrangian of QCD instead of the effective chiral Lagrangian, 
one may write the normalized partition function as
\begin{align}
  Z(M)
  &=\int [\mathcal{D}A]\ {\prod_{n}}'\det(\lambda_n^2+M^\dagger
  M)\ \ee^{-S_g}\bigg/ \int [\mathcal{D}A]\
  \Big({\prod_{n}}'\lambda_n^2\Big)^{N_f}\ \ee^{-S_g}\\
  &= \biggl\langle{\prod_{n}}'\det\left(1+\frac{M^\dagger
  M}{\lambda_n^2}\right)\biggr\rangle\,,
  \label{eq:Z_micro}
\end{align}
where $\displaystyle S_g\equiv\int d^4x\,F^a_{\mu\nu}F^a_{\mu\nu}/4$ and 
$\langle\cdots\rangle$ represents expectation values with respect to the measure in the chiral limit. 
$\prod'_n$ (and later $\sum'_n$) denotes the product (sum) over all eigenvalues with $\re{\lambda_n>0}$. 
We neglect the anomaly which is suppressed at large $\mu$, and thus we assume that no zero modes appear. Equating (\ref{eq:Z_micro}) with (\ref{eq:Z_macro}) 
for $N_f\geq 4$ and matching the coefficients at $O(M^2)$ and $O(M^4)$, we arrive at novel spectral sum rules,
\beq
  \biggl\langle{\sum_{n}}'\frac{1}{\lambda_n^2}\biggr\rangle
  =\biggl\langle{\sum_{m<n}}'\frac{1}{\lambda_m^2\lambda_n^2}\biggr\rangle
  =\biggl\langle{\sum_{n}}'\frac{1}{\lambda_n^6}\biggr\rangle=0\,,
  \qquad 
  \biggl\langle{\sum_{n}}'\frac{1}{\lambda_n^4}\biggr\rangle
  =\frac{9}{4\pi^4(N_f-1)^2}(V_4\Delta^2)^2\,.
  \label{eq:sumrules}
\eeq
The vanishing of many of the spectral sums is a salient feature of the high-density limit by which 
(\ref{eq:sumrules}) is distinguished most clearly from the conventional 
spectral sum rules at $\mu=0$ \cite{Leutwyler1992,Smilga1995}. We add that matching between 
(\ref{eq:Z_micro}) and (\ref{eq:Z_macro_2}) for $N_f=2$ reveals that (\ref{eq:sumrules}) is correct 
for $N_f=2$ as well.

Introducing the spectral density $\rho(\lambda)$ and the microscopic spectral density $\rho_s(\lambda)$ defined by
\beq
  \rho(\lambda)\equiv \biggl\langle\sum_n\delta^2(\lambda-\lambda_n)\biggr\rangle \qquad \text{and}\qquad 
  \rho_s(z)\equiv\lim_{V_4\to\infty}\frac{\pi^2}{3V_4\Delta^2}\,
  \rho\biggl(\frac{\pi z}{\sqrt{3V_4\Delta^2}}\biggr)\,,
  \label{eq:rho_s}
\eeq
the second sum rule in (\ref{eq:sumrules}) is rewritten as
\beq
  \int\limits_{\text{Re\,}z>0}   \hspace{-5pt}   d^2z~\frac{\rho_s(z)}{z^4}
  = \frac{1}{4(N_f-1)^2}\,.
\eeq
In analogy with the $\mu=0$ case, this formula strongly suggests that
$\rho_s(z)$ is a universal function determined by the global
symmetries of the problem. It leads to the  
observation that the smallest eigenvalues of $\D$ sit at the scale of $O(1/\sqrt{V_4})$ 
and that their distribution is governed by $\Delta$, in contrast to the situation at $\mu=0$ where the magnitude 
of the smallest eigenvalues is $O(1/V_4)$ and the quantity relevant for level correlations is not $\Delta$, but $\Sigma$. 

Finally we comment on the generalization of the sum rules (\ref{eq:sumrules}) to the massive case. It is 
achieved by rescaling both the eigenvalues and the masses simultaneously as the volume is taken to infinity. We begin with the simplest 
case, i.e., $N_f=2$ with masses $m_1$ and $m_2$. In terms of rescaled dimensionless variables,
\beq
  z_n\equiv \lambda_n \frac{\sqrt{3V_4\Delta^2}}{\pi}\,, \qquad \m_i \equiv m_i \frac{\sqrt{3V_4\Delta^2}}{\pi}\,,
\eeq
the simplest sum rule reads
\beq
  \alangle{\sum_n}'\frac{1}{z_n^2+\m_1^2}\arangle
  =\frac{\m_2^2}4\,\frac{I_0(x)-I_2(x)}{I_0(x)}\qquad\text{with}\quad
  x=\m_1\m_2\,,
  \label{eq:massive_sum_rule}
\eeq
where $\langle\!\langle\cdots\rangle\!\rangle$ represents expectation values with respect to the massive measure. 
For larger $N_f$ the explicit expressions become increasingly involved. For $N_f=4$ with equal masses, we have
\beq
  \alangle{\sum_n}'\frac{1}{z_n^2+\m^2}\arangle=\frac{2I_0(y)I_1(y)-3I_1(y)I_2(y)+I_2(y)I_3(y)}
  {4(3I_0(y)^2-4I_1(y)^2+3I_2(y)^2)}
  \qquad\text{with}\quad y=\m^2\,.
\eeq
Similarly we can generalize $\rho_s$ in (\ref{eq:rho_s}) to the double-microscopic spectral density defined by
\beq
\label{eq:rho_sNf}
  \rho_s^{(N_f)}(z;\m_1,\dots,\m_{N_f})\equiv\lim_{V_4\to\infty}
  \frac{\pi^2}{3V_4\Delta^2}\rho\bigg(\frac{\pi z}{\sqrt{3V_4\Delta^2}}\bigg)\bigg|
  _{\m_i=m_i\frac{\sqrt{3V_4\Delta^2}}\pi\text{ fixed}}\,,
\eeq
in terms of which we can rewrite, e.g., (\ref{eq:massive_sum_rule}) in
the form
\beq
  \int\limits_{\text{Re\,}z>0}   \hspace{-5pt}   d^2z~\frac{\rho_s^{(2)}(z;\m_1,\m_2)}{z^2+\m_1^2}
  =\frac{\m_2^2}{4}\frac{I_0(x)-I_2(x)}{I_0(x)}\,.
\eeq

As emphasized in the introduction, it has been firmly established at $\mu=0$ by various arguments %
\cite{Verbaarschot:2000dy} that the functions $\rho_s$ and $\rho_s^{(N_f)}$ are universal in the sense that they depend solely on 
the pattern of spontaneous symmetry breaking and not on the detailed form of the microscopic interactions. Hence we 
expect that the same holds at large $\mu$ as well, even though the
explicit forms of $\rho_s$ and $\rho_s^{(N_f)}$ defined in (\ref{eq:rho_s}) and (\ref{eq:rho_sNf})
are still unknown. A promising approach will be to construct an appropriate ChRMT 
that corresponds to dense two-color QCD (work in progress). In this regard one might be tempted to consider a
minimal modification of the conventional ChRMTs by addition of a $\gamma_0\mu$-term, but this does not suffice, since 
in the large-$\mu$ limit the latter will dominate the Dirac operator completely 
and render its dynamics trivial.

\section{Summary}   \label{5}

We have constructed the low-energy effective Lagrangian 
for dense two-color QCD with an even number of flavors at large quark
chemical potential $\mu$ based on the symmetry breaking pattern
induced by the formation of a diquark condensate \cite{Kanazawa:2009ks}. 
Also, we have identified a new finite-volume $\epsilon$-regime for the superfluid phase
at large $\mu$.  In this regime, we can
exactly determine the quark mass dependence of the partition function
from the effective theory.  Matching this result against the two-color
QCD partition function, we have derived novel spectral sum rules for
inverse powers of the complex eigenvalues of the Dirac operator.  Our
sum rules explicitly show that the Dirac spectrum at large $\mu$ is
governed by the fermion gap $\Delta$, unlike the spectrum at low
$\mu$, which is governed by the chiral condensate as shown in
\cite{Smilga1995}.

  Since the fermion sign problem is absent in this theory, 
our sum rules can in principle be checked in lattice QCD simulations.  
This is in contrast to real (three-color) QCD where the severe sign problem prevents us from
observing the presumed color superconductivity, although similar
spectral sum rules could be derived in the corresponding
$\epsilon$-regime \cite{Yamamoto:2009ey}.  Testing our sum rules for two
colors on the lattice would enable us to measure the value of the BCS gap $\Delta$ at large $\mu$ 
for the first time, since previous studies of two-color
QCD at nonzero $\mu$ have only been able to determine the magnitude of
the diquark condensate, not the gap itself.

It would be interesting to
obtain an analogue of the Banks-Casher relation at large $\mu$ and the
concrete form of the microscopic spectral density in the
$\epsilon$-regime we identified.  In particular, it is a challenging
problem to construct the corresponding random matrix model, which has
turned out to be very successful at $\mu=0$ and small $\mu$, to
reproduce our spectral sum rules and to elucidate universal properties
of the Dirac spectrum at large $\mu$ for both two and three colors.
A detailed analysis of these issues is deferred to future work.

\acknowledgments

TK and NY are supported by the Japan Society for the Promotion of
Science for Young Scientists.  TW is supported by DFG and acknowledges
additional support by JSPS and by the G-COE program of the University
of Tokyo.

\bibliographystyle{JHEP}
\bibliography{Lattice09}

\providecommand{\href}[2]{#2}\begingroup\raggedright\begin{thebibliography}{10}

\bibitem{Rajagopal2000}
K.~Rajagopal and F.~Wilczek, {\it {The condensed matter physics of QCD}},
  \href{http://arxiv.org/abs/hep-ph/0011333}{{\tt hep-ph/0011333}}.

\bibitem{Alford2008}
M.~G. Alford, A.~Schmitt, K.~Rajagopal, and T.~Sch\"afer, {\it {Color
  superconductivity in dense quark matter}},  {\em Rev. Mod. Phys.} {\bf 80}
  (2008) 1455--1515, [\href{http://arxiv.org/abs/0709.4635}{{\tt
  arXiv:0709.4635}}].

\bibitem{Alford:1998mk}
M.~G. Alford, K.~Rajagopal, and F.~Wilczek, {\it {Color-flavor locking and
  chiral symmetry breaking in high density {QCD}}},  {\em Nucl. Phys.} {\bf
  B537} (1999) 443--458, [\href{http://arxiv.org/abs/hep-ph/9804403}{{\tt
  hep-ph/9804403}}].

\bibitem{Leutwyler1992}
H.~Leutwyler and A.~V. Smilga, {\it {Spectrum of Dirac operator and role of
  winding number in QCD}},  {\em Phys. Rev.} {\bf D46} (1992) 5607--5632.

\bibitem{Shuryak:1992pi}
E.~V. Shuryak and J.~J.~M. Verbaarschot, {\it {Random matrix theory and
  spectral sum rules for the Dirac operator in QCD}},  {\em Nucl. Phys.} {\bf
  A560} (1993) 306--320, [\href{http://arxiv.org/abs/hep-th/9212088}{{\tt
  hep-th/9212088}}].

\bibitem{Verbaarschot:2000dy}
J.~J.~M. Verbaarschot and T.~Wettig, {\it {Random matrix theory and chiral
  symmetry in QCD}},  {\em Ann. Rev. Nucl. Part. Sci.} {\bf 50} (2000)
  343--410, [\href{http://arxiv.org/abs/hep-ph/0003017}{{\tt hep-ph/0003017}}].

\bibitem{Fukaya:2007fb}
{\bf JLQCD} Collaboration, H.~Fukaya {\em et~al.}, {\it {Two-flavor lattice QCD
  simulation in the epsilon-regime with exact chiral symmetry}},  {\em Phys.
  Rev. Lett.} {\bf 98} (2007) 172001,
  [\href{http://arxiv.org/abs/hep-lat/0702003}{{\tt hep-lat/0702003}}].

\bibitem{Splittorff:2008sw}
K.~Splittorff and J.~J.~M. Verbaarschot, {\it {Lessons from Random Matrix
  Theory for QCD at Finite Density}},
  \href{http://arxiv.org/abs/0809.4503}{{\tt arXiv:0809.4503}}.

\bibitem{Yamamoto:2009ey}
N.~Yamamoto and T.~Kanazawa, {\it {Dense QCD in a Finite Volume}},  {\em Phys.
  Rev. Lett.} {\bf 103} (2009) 032001,
  [\href{http://arxiv.org/abs/0902.4533}{{\tt arXiv:0902.4533}}].

\bibitem{Kogut2000}
J.~B. Kogut, M.~A. Stephanov, D.~Toublan, J.~J.~M. Verbaarschot, and
  A.~Zhitnitsky, {\it {QCD-like theories at finite baryon density}},  {\em
  Nucl. Phys.} {\bf B582} (2000) 477--513,
  [\href{http://arxiv.org/abs/hep-ph/0001171}{{\tt hep-ph/0001171}}].

\bibitem{Kanazawa:2009ks}
T.~Kanazawa, T.~Wettig, and N.~Yamamoto, {\it {Chiral Lagrangian and spectral
  sum rules for dense two- color QCD}},  {\em JHEP} {\bf 08} (2009) 003,
  [\href{http://arxiv.org/abs/0906.3579}{{\tt arXiv:0906.3579}}].

\bibitem{Schafer:1998ef}
T.~Sch\"afer and F.~Wilczek, {\it {Continuity of quark and hadron matter}},
  {\em Phys. Rev. Lett.} {\bf 82} (1999) 3956--3959,
  [\href{http://arxiv.org/abs/hep-ph/9811473}{{\tt hep-ph/9811473}}].

\bibitem{Hatsuda:2006ps}
T.~Hatsuda, M.~Tachibana, N.~Yamamoto, and G.~Baym, {\it {New critical point
  induced by the axial anomaly in dense QCD}},  {\em Phys. Rev. Lett.} {\bf 97}
  (2006) 122001, [\href{http://arxiv.org/abs/hep-ph/0605018}{{\tt
  hep-ph/0605018}}].

\bibitem{Gasser:1987ah}
J.~Gasser and H.~Leutwyler, {\it {Thermodynamics of Chiral Symmetry}},  {\em
  Phys. Lett.} {\bf B188} (1987) 477.

\bibitem{Smilga1995}
A.~V. Smilga and J.~J.~M. Verbaarschot, {\it {Spectral sum rules and finite
  volume partition function in gauge theories with real and pseudoreal
  fermions}},  {\em Phys. Rev.} {\bf D51} (1995) 829--837,
  [\href{http://arxiv.org/abs/hep-th/9404031}{{\tt hep-th/9404031}}].

\end{thebibliography}\endgroup
\end{document}